# Stimulated Raman Scattering and Nonlinear Focusing of High-Power Laser Beams Propagating in Water

B. Hafizi, J.P. Palastro*, J.R. Peñano, D.F. Gordon, T.G. Jones, M.H. Helle and D. Kaganovich

*Naval Research Laboratory, Washington DC 20375*

*Icarus Research, Inc, PO Box 30780, Bethesda MD 20824-0780*

**Abstract**

The physical processes associated with propagation of a high-power (power > critical power for self-focusing) laser beam in water include nonlinear focusing, stimulated Raman scattering (SRS), optical breakdown and plasma formation. The interplay between nonlinear focusing and SRS is analyzed for cases where a significant portion of the pump power is channeled into the Stokes wave. Propagation simulations and an analytical model demonstrate that the Stokes wave can re-focus the pump wave after the power in the latter falls below the critical power. It is shown that this novel focusing mechanism is distinct from cross-phase focusing. While discussed here in the context of propagation in water, the gain-focusing phenomenon is general to any medium supporting nonlinear focusing and stimulated forward Raman scattering.

Several physical processes accompany the propagation of high-power laser beams in water [1-7], including nonlinear focusing, stimulated Raman scattering (SRS) and ionization. The interplay between these leads to complex spatio-temporal evolution of the beam. For instance, during self-focusing a high power laser beam can undergo SRS transferring most of its energy to the Stokes wave [5]. The purpose of this paper is to point out, for the first time, that the simultaneous occurrence of self-focusing and SRS results in gain-focusing, or a refocusing of the pump wave by Stokes wave after the pump has been strongly depleted by the Stokes wave. Self-consistent propagation simulations and an analytical model, both incorporating the key physical processes, demonstrate the effect and distinguish it from cross-phase focusing [8]. While discussed here in the context of propagation in water, the gain-focusing phenomenon is general to any medium supporting nonlinear focusing and stimulated forward Raman scattering.

For simplicity the presentation is limited to the case where the anti-Stokes component is not appreciably excited [5] and the electric field can be written as $E(\mathbf{r},t) = \frac{1}{2} E_P e^{i\varphi_P} + \frac{1}{2} E_S e^{i\varphi_S} + c.c.$, where $\phi_j \equiv k_j z - \omega_j t$ is the phase, the wavenumber $k_j$ and frequency $\omega_j$ are related by the refractive index $n = c k_j / \omega_j$, $E_j$ is the amplitude ($j = P, S$) and the suffices $P$ and $S$ refer to the pump and Stokes components, respectively. The Raman scattering process involves stretching of $\text{O-H}$ bonds in $\text{H}_2\text{O}$ molecules [8] (vibration frequency $\omega_v \approx 3450 \, \text{cm}^{-1}$), driven by a force that is proportional to the time-average of the electric field squared. The associated non-linear polarization density depends on bond stretching and therefore on $<E(\mathbf{r},t)^2> = \frac{1}{2} E_P E_S^* e^{i(\varphi_P - \varphi_S)} + c.c.$ Extending the analyses in Refs. [8] and [9] results in the following propagation equation

$$\left( \nabla_\perp^2 + 2ik_S \frac{\partial}{\partial z} \right) E_S = -\frac{6\pi \omega_S^2}{c^2} [(\chi_R + \chi_{NR}) |E_P|^2 + \tfrac{1}{2} \chi_{NR} |E_S|^2] E_S + \left[ \frac{\omega_p^2}{c^2} - \frac{8\pi i \omega_S}{c^2} \frac{U_{ion}}{|E_S|^2} \left( \frac{\partial N_e}{\partial \tau} \right)_{S, MPI} \right] E_S \qquad (1)$$

for the Stokes wave; the pump pulse obeys a similar equation with $\chi_R \to \chi_R^*$ for the Raman susceptibility [8] along with suffix interchange. By including the full equations for both the pump and Stokes waves, this model avoids making the "strong-pump" approximation. In Eq. (1)

$\nabla_\perp^2$ is the transverse Laplacian operator, $\chi_{NR}$ is a real-valued non-resonant (Kerr-type) susceptibility [8,9], the-next-before last term on the RHS represents the effects of plasma with frequency $\omega_p = (4\pi N_e e^2/m)^{1/2}$ and free electron density $N_e(r,z,\tau)$ [10]. Here resonant stimulated Raman scattering is considered such that $\chi_R$ is purely imaginary [8]. The imaginary component of the susceptibility arises from population loss of the laser-excited vibrational state through interactions with the external environment. As a result, the Raman response can be in phase (as opposed to in quadrature) with the pump (Stokes) field allowing it to transfer energy to the Stokes (pump) wave. The loss of photons in the process of multiphoton ionization is represented by the last term on the RHS, where $U_{ion}$ is the ionization potential of water molecules and the suffix $S$ on the time derivative indicates the change in density due to absorption of Stokes photons. Finite pulse-length and group velocity dispersion effects [11] are found to be negligible for parameters of interest.

Free electrons are generated through multiphoton ionization with the number density evolving according to $\partial N_e/\partial \tau = W_{MPI} N_{H_2O}$, where $W_{MPI}$ is the multi-photon ionization rate, $\tau = t - z/v_g$ is the time-variable in the laser pulse frame and $v_g$ is the linear group velocity [10]. The multi-photon ionization rate is $W_{MPI} = 2\pi \sum_j \omega_j [I_j(r,z,\tau)/I_{MPI}]^{\ell_j}/(\ell_j - 1)!$, where $j = P, S$, $\ell_j = Int[U_{ion}/\hbar\omega_j + 1]$ is an integer denoting the number of photons required for ionization, $\hbar\omega$ is the photon energy, $I_j(r,z,\tau)$ is the intensity and $I_{MPI}$ is the characteristic ionization intensity. It is assumed that any given multiphoton ionization event is effected by pump photons or by Stokes photons independently. For the cases considered in this paper, simulations (presented below) justify the neglect of inverse Bremsstahlung, attachment and recombination [12] and avalanche ionization. Additionally it is assumed that hydrated electrons [13] do not fundamentally alter the physics, because the pump wavelength is not strongly resonant.

Propagation simulations [14-16] solving Eq. (1) in axi-symmetric, cylindrical coordinates (and a similar equation for the pump pulse) along with the electron density equation were performed employing the parameters listed in Table 1, which are similar to those of experiments presented in Ref. 5. Figure 1 displays the normalized fluence as a function of transverse coordinate and propagation distance for the pump (top) and Stokes (bottom) pulses. The dashed red line demarcates the radius containing 25% of the power in each pulse. The red solid line is a magnified version of the 25% radius curve to illustrate the transverse dynamics more clearly. During the initial lens focusing the pump pulse transfers energy to the Stokes pulse. The energy transfer occurs more rapidly where the pump intensity is large, providing a gain narrowed Stokes profile reaching a minimum spot radius at ~10 cm. At this point, most of the energy is contained in the Stokes pulse, which having reached a high intensity ionizes the water. The refraction of both pulses from the ionized electrons can be observed in the rapid increase in the spot radii and the gap in the pump pulse fluence.

In the absence of the Stokes pulse, the pump pulse, now being well below the critical power for self-focusing, would diffract. What is observed, however, is a re-focusing of the pump pulse. This re-focusing could potentially be explained by the cross-phase modulation [8] between the pump and Stokes pulses mediated by the real susceptibility $\chi_{NR}$. However, $\chi_{NR} < |\chi_R|$ and the refocusing persists in simulations (not shown) when $\chi_{NR}$ is set explicitly equal to zero. This refocusing results from gain-focusing of the pump pulse by the Stokes pulse. In the presence of an imaginary susceptibility, $\mathrm{Im}\,\chi_R \neq 0$, that couples the two beams, one pulse can focus the other. This is demonstrated more clearly with an analytical model.

To analyze the nonlinear propagation use is made of the source-dependent expansion (SDE) method in which self-similar forms for the complex-valued amplitudes are substituted into the propagation equations [10]. A novel extension of the SDE method is applied, employing two envelopes—one each for the pump and Stokes. The SDE method uses the ansatz that the field amplitudes are peaked on axis and have Gaussian forms $E_j(r,z) = B_j(z)\exp\{i\theta_j(z) - [1 + i\alpha_j(z)]r^2/R_j^2(z)\}$, where $B_j$ is the amplitude, $\theta_j$ is the phase, $R_j$ is the spot radius and $\alpha_j$ is related to wavefront curvature for the $j-$th wave. An additional approximation in the analysis is that the same number of pump and Stokes photons are required to ionize a water molecule.

Following the SDE method, the variation of spot radius is given by [10]

$$\frac{\partial^2 R_j}{\partial z^2} + \frac{1}{R_j}\frac{\partial}{\partial z}(R_j^2 G_j'') + \frac{4}{k_j^2 R_j^3}\left[-1 - k_j R_j^2 G_j' + \tfrac{1}{4}(k_j R_j^2 G_j'')^2\right] = 0. \tag{2}$$

where $G_j \equiv G_j' + i G_j''$. The $-1$ term in square brackets is associated with vacuum diffraction of electromagnetic beams. The full expression for $G_j$ is unwieldy; here it suffices to present only the terms relevant to nonlinear focusing: $k_s R_S^2 G_S' = -P_S/P_{Sc} - 8(\omega_S/\omega_P)^2 P_P/P_{Pc}/(R_P^2/R_S^2+1)^2$, $k_s R_S^2 G_S'' = -8(\omega_S/\omega_P)^2 (\chi_R/\chi_{NR}) P_P/P_{Pc}/(R_P^2/R_S^2+1)^2$ for Stokes wave and $k_P R_P^2 G_P' = -P_P/P_{Pc} - 8 P_S/P_{Pc}/(R_S^2/R_P^2+1)^2$, $k_P R_P^2 G_P'' = 8(\chi_R/\chi_{NR}) P_S/P_{Pc}/(R_S^2/R_P^2+1)^2$ for the pump wave. In these expressions, $P_j$ is the power, and $P_{jc}$ is the critical power for self-focusing. The first term in the expression for $G_j'$ is associated with self-focusing while the second term is associated with cross-beam focusing [8]. These represent refractive focusing. The pure imaginary Raman susceptibility $\chi_R$ only appears in $G_j''$. In particular, the term $+\tfrac{1}{4}(k_j R_j^2 G_j'')^2$ in Eq. (2) is positive definite and always focusing. This results in gain focusing. The utility of the SDE approach is that Eq. (2) shows explicitly the imaginary susceptibility, through $G_j''$, and the real susceptibility, through $G_j'$, result in distinct spot size dynamics.

Figure 2 (a) plots pump and Stokes powers as a function of propagation distance for the same parameters as the simulations (Table 1). By $\sim 30$ cm the pump pulse is strongly depleted and would be unable to self-focus. Figure 2 (b) shows the variation of the pump and Stokes spot radii indicating that the former re-focuses as mentioned in connection with Fig. 1. Figure 3 displays the same quantities when the imaginary part of the Raman susceptibility in the expression for $G_P''$ is neglected ($\chi_R \to 0$). In this case, the pump spot does not refocus, demonstrating that $\chi_R$ is responsible for the refocusing.

In conclusion, both simulations and an analytical model for high-power laser propagation in water demonstrate the distinct focusing contributions of the real and the imaginary parts of the Raman susceptibility. The *real* part contributes to refractive focusing while the *imaginary* part leads to gain focusing. While discussed here in the context of propagation in water, the gain-focusing phenomenon is general to any medium supporting nonlinear focusing and stimulated forward Raman scattering.

| | |
|---|---|
| Self-Focusing Power (@ 532 nm) $P_{Pc}$ | 1.7 MW[a] |
| Pump Wavelength $\lambda_P$ | 532 nm |
| Pump Power at Input $P_P$ | 2.5 $P_{Pc}$ |
| Initial Pump Spot Radius $R_P$ | 0.375 mm |
| Initial Pump Duration (flat top) | 100 ps |
| Initial Pump Spot Divergence | -0.6 mrad |
| Stokes Wavelength $\lambda_S$ | 652 nm |
| Initial Stokes Power $P_S$ | 2.21x10$^{-5}$ W [b] |
| Initial Stokes Spot Radius $R_S$ | 0.375 mm |
| Refractive Index $n$ | 1.334 |
| Stokes Raman Susceptibility $\chi_R$ | $-i$ 8.3x10$^{-15}$ cm$^3$/erg [c] |
| Ratio of Susceptibilities $-\operatorname{Im}\chi_R(\omega_S)/\chi_{NR}$ | 4.4 |
| Ionization Potential of Water $U_{ion}$ | 8.5 eV [d] |
| Multiphoton Ionization Intensity $I_{MPI}$ | 4x10$^{13}$ W/cm$^2$ |

Table 1. Parameters for simulations of stimulated Raman scattering in water.

[a] From Ref. 17

[b] From Ref. 1

[c] Adjusted to match Raman growth rate quoted in Ref. 1.

[d] Compiled from Refs. 18 & 19.

**Acknowledgements**

The authors would like to acknowledge helpful discussions with Y.-H. Chen and A. Ting. This work was supported by the Naval Research Laboratory Base Program.

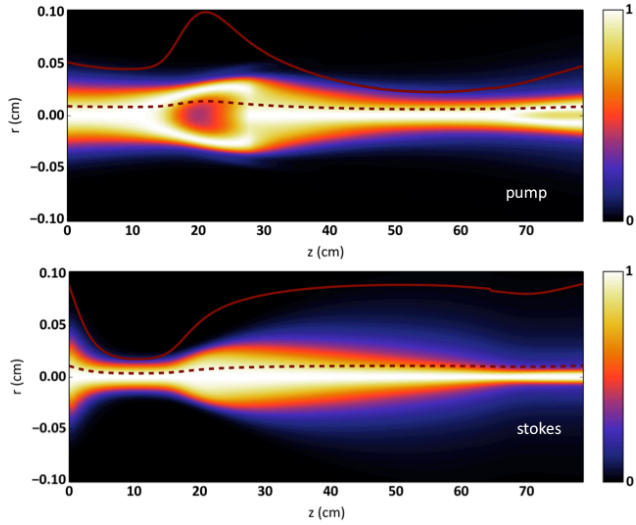

Figure 1. Energy fluence normalized to the maximum at each axial position for the pump (top) and Stokes (bottom) pulses as a function of transverse coordinate and propagation distance. The dashed red line traces the radius containing 25% of the power in the pump and Stokes pulses. The solid red line is a magnified version of the dashed line for clarity. By 20 cm the pump is almost completely depleted but refocuses through its interaction with the Stokes pulse mediated by the imaginary component of the resonant susceptibility. This is gain-focusing.

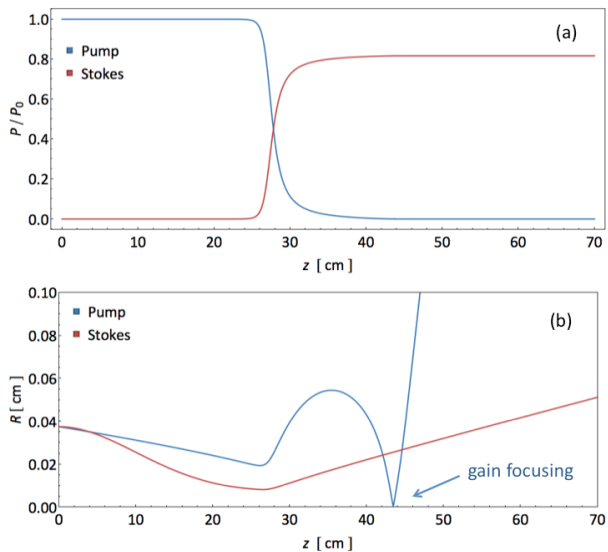

Figure 2. (a) normalized Pump (blue) and Stokes (red) power versus propagation distance based on the analytical source-dependent expansion approach. (b) corresponding Pump and Stokes spot radii versus propagation distance.

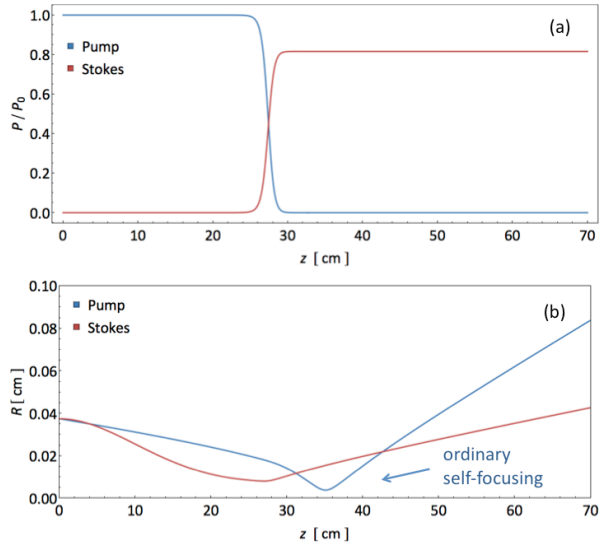

Figure 3. (a) normalized Pump (blue) and Stokes (red) power versus propagation distance based on the analytical source-dependent expansion approach with the imaginary part of the Raman susceptibility nulled. (b) corresponding Pump and Stokes spot radii versus propagation distance.